\documentclass[%
 reprint,
superscriptaddress,
longbibliography,
showkeys,
 amsmath,amssymb,
 aps,
pre, 
]{revtex4-2}

\usepackage{graphicx}
\usepackage[caption=false]{subfig} 

\usepackage{float}
     
\usepackage{dcolumn}
\usepackage{bm}

\newcommand{\ee}{\end{equation}} 
\newcommand{\be}{\begin{equation}}

\usepackage[mathscr]{euscript}  
\usepackage{xcolor}  

\usepackage{tcolorbox}
\usepackage{comment}
\usepackage{float}

\makeatletter
\newsavebox{\@brx}
\newcommand{\llangle}[1][]{\savebox{\@brx}{\(\m@th{#1\langle}\)}%
  \mathopen{\copy\@brx\kern-0.5\wd\@brx\usebox{\@brx}}}
\newcommand{\rrangle}[1][]{\savebox{\@brx}{\(\m@th{#1\rangle}\)}%
  \mathclose{\copy\@brx\kern-0.5\wd\@brx\usebox{\@brx}}}
\makeatother

\begin{document} 

\preprint{APS/123-QED}

\title{Optimal area exploration by resetting active particles}

\author{Kristian St\o{}levik Olsen}
\email{kristian.olsen@hhu.de}
\affiliation{Institut für Theoretische Physik II - Weiche Materie, Heinrich-Heine-Universität Düsseldorf, D-40225 Düsseldorf, Germany}


\author{Hartmut L\"{o}wen}
\affiliation{Institut für Theoretische Physik II - Weiche Materie, Heinrich-Heine-Universität Düsseldorf, D-40225 Düsseldorf, Germany}

\author{Lorenzo Caprini}
\affiliation{Physics Department, Sapienza University of Rome, P.le Aldo Moro 5, IT-00185, Rome, Italy}
\email{lorenzo.caprini@gssi.it, lorenzo.caprini@uniroma1.it}

\begin{abstract}
Identifying optimal strategies for efficient spatial exploration is crucial, both for animals seeking food and for robotic search processes, where maximizing the covered area is a fundamental requirement. Here, we propose position resetting as an optimal protocol to enhance spatial exploration in active matter systems. Specifically, we show that the area covered by an active Brownian particle exhibits a non-monotonic dependence on the resetting rate, demonstrating that resetting can optimize spatial exploration.
Our results are based on experiments with active granular particles undergoing Poissonian resetting and are supported by active Brownian dynamics simulations. The covered area is analytically predicted at both large and small resetting rates, resulting in a scaling relation between the optimal resetting rate and the self-propulsion speed.
\end{abstract}

\pacs{Valid PACS appear here} 
\maketitle

\noindent

\noindent

Effective navigation and exploration of an environment is a crucial task for a myriad of systems, spanning from the microscale to the macroscale. Specifically, this need is encountered in many active and living systems, which convert energy from the environment into directed motion~\cite{marchetti2013hydrodynamics, elgeti2015physics, bechinger2016active}. 
In the realm of synthetic active matter, such as robotics, efficient exploration has a wide range of applications, including mapping uncharted territories and conducting life-saving rescue missions~\cite{yamauchi1998frontier,burgard2000collaborative,albers2002exploring}. The need for efficient spatial exploration also extends to bio-inspired technologies, such as micro-robotic drug delivery and cargo transport~\cite{naahidi2013biocompatibility,palagi2018bioinspired}.
Likewise, for living organisms, the constant quest for nutrients and other resources is an essential task to survive~\cite{viswanathan2011physics,volpe2017topography}.  
In this broad range of systems, an agent is more likely to achieve its goal when the spatial exploration is efficient, i.e.,\ when the covered area is maximized during the search process.

\begin{figure}[t!]
    \centering
    \includegraphics[width = 0.9\columnwidth]{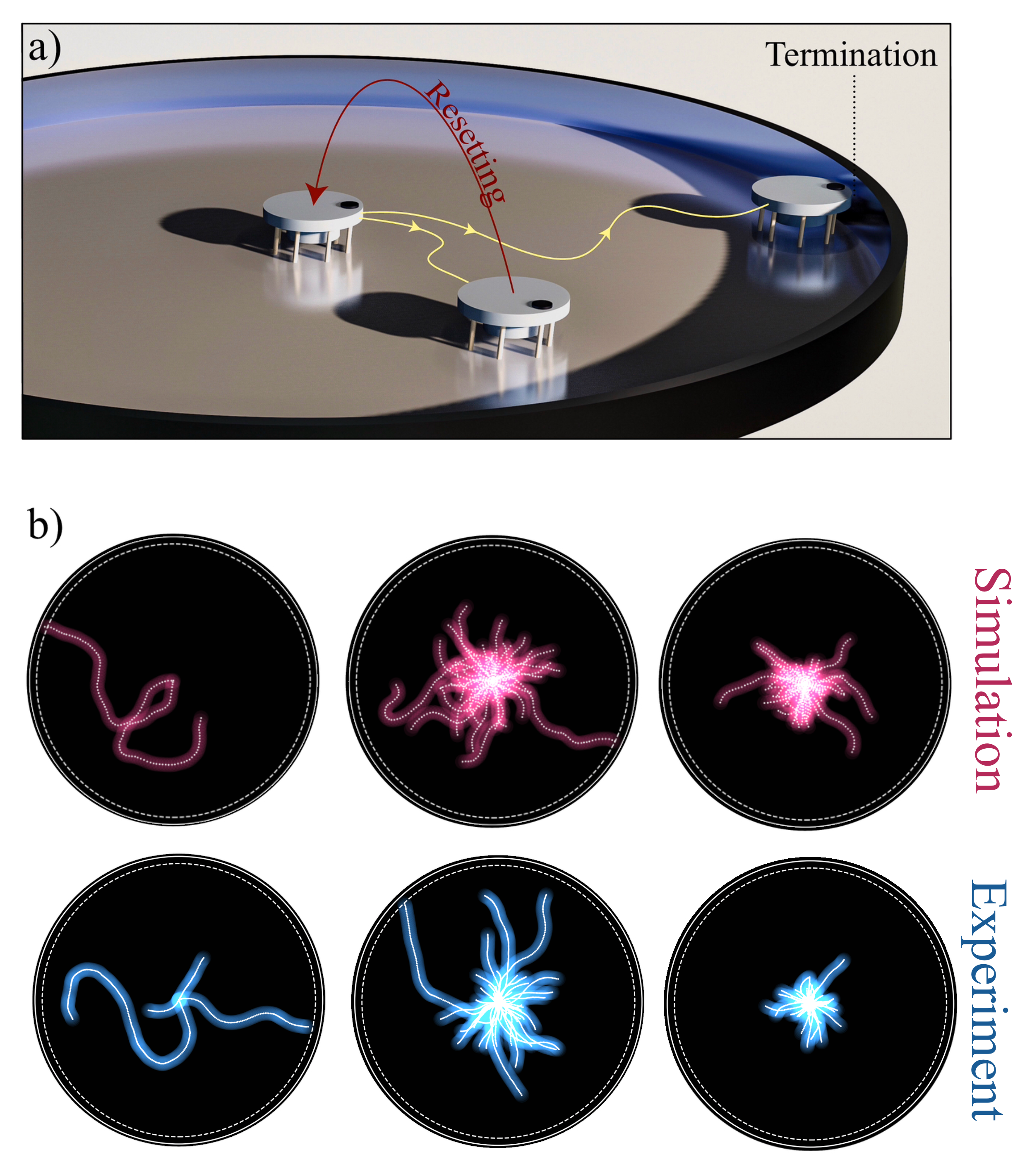}
    \caption{\textbf{Active granular particle under resetting.}
    (a) Visualization of the experimental setup, with a vibrobot moving in a circular arena. The particle is reset (red arrow) back to its initial position with orientation randomization at random times extracted from an exponential distribution. This process lasts until the particle hits the boundary of the plate. 
    (b)-(c) Time-trajectories of finite-sized active particles in a circular plate with radial size $R$ for three different values of the resetting rate, $r=0.5,1.0,2.5 \text{ Hz}$. The covered area is obtained by coloring the center-of-mass trajectory with a thickness corresponding to the particle diameter. (b) is obtained from experiments, while (c) by simulations. 
    These typical trajectories qualitatively reveal a non-monotonic covered area with the resetting rate $r$, both in experiments and simulations. 
    }
    \label{fig:paths}
\end{figure}

Recently, resetting problems~\cite{evans2020stochastic,gupta2022stochastic} have stimulated research in non-equilibrium statistical physics and stochastic processes, with a surge of interest after the seminal work by Evans and Majumdar more than a decade ago~\cite{evans2011diffusion,evans2011diffusion2}. Resetting consists of intermittent returns to a fixed location, and produces non-equilibrium steady states with non-trivial relaxation properties \cite{majumdar2015dynamical,gupta2021resetting,goerlich2024resetting,boyer2024power}, as well as a high degree of entropy production \cite{fuchs2016stochastic,busiello2020entropy,mori2023entropy,olsen2024thermodynamic,olsen2024thermodynamic2,gupta2022work}. A large proportion of studies focus on first passage properties under resetting, which have shown the potential to mitigate long search times~\cite{pal2019first,pal2017first,faisant2021optimal,ahmad2019first,besga2020optimal,biroli2024resetting,ahmad2019first,mercado2022reducing,mendez2024first,bressloff2020search}. The first passage time, i.e.\ the minimal time to approach a target, can in many cases be minimized by tuning the resetting rate. This strategy cuts off unfavorable, long trajectories, which do not encounter the target, while favorable trajectories find the target without strong influences due to resetting~\cite{pal2017first}. 
The theoretical picture developed has also been validated experimentally, for instance by considering colloids in optical tweezers~\cite{faisant2021optimal,tal2020experimental,goerlich2023experimental}. 
In addition, beyond the first-passage time, several theoretical studies have focused on other functionals~\cite{meerson2023geometrical,hartmann2024first,singh2022first} under resetting, as well as on convex-hull problems~\cite{mukherjee2024large}.

Beyond the well-known passive scenario, increasing attention has been devoted to non-equilibrium systems, such as active matter, where
resetting has been studied rather intensively from a theoretical perspective. Several recent results have been obtained for one-dimensional run-and-tumble particles~\cite{evans2018run,iyaniwura2024asymptotic,bressloff2020occupation, hartmann2020convex,singh2022mean,radice2021one,tucci2022first,bressloff2021accumulation,linn2023first,di2023current, KSO2023} as well in higher dimensions with both run and tumble~\cite{santra2020run,smith2022exact,hartmann2020convex} and active Brownian dynamics~\cite{kumar2020active,abdoli2021stochastic,pal2024activethermal,Baouche_2024,Baouche2025OptimalFT} even in confining geometries~\cite{sar2023resetting}. Here, experimental studies are sparse, except for recent works using hexbugs and programmable robots~\cite{altshuler2023environmental,PRXLife.2.033007}.

However, while it is known that resetting affects the first passage time, its role in the covered area problem is not clear. 
Intuitively, once a region has been mapped, there is no benefit in revisiting. 
From this point of view, resetting could naively represent a bad strategy, since it forces the active particle to re-examine already visited locations. 
However, resets also facilitate a dense exploration of a neighboring region of the resetting location, a feature that is likely lost in the absence of resets in favor of sparse exploration.

Here, we discover that resetting provides an efficient strategy to enhance the area covered by active matter systems.
In particular, we experimentally and numerically study the dynamics of an active particle confined on a circular plate and subject to stochastic resetting. This study is performed by considering an active vibrobot~\cite{scholz2018inertial,antonov2024inertial} whose position is reset to the middle of the plate after a random time, which is extracted from an exponential distribution. At each reset, the particle orientation is randomized. The exploration process ends once the boundary of the system is reached (Fig.~\ref{fig:paths}~a).
We demonstrate the existence of an activity-dependent optimal resetting rate that maximizes the covered area, as visible from direct inspection of particle trajectories for a vibrobot (Fig.~\ref{fig:paths}~b) obtained both experimentally and numerically.
Our study implies that resetting can be adopted as an optimal strategy to enhance spatial exploration in robotics applications, for instance to enhance the efficiency during the search of targets, such as food and resources.

\begin{figure}
    \centering
    \includegraphics[width=7.1cm]{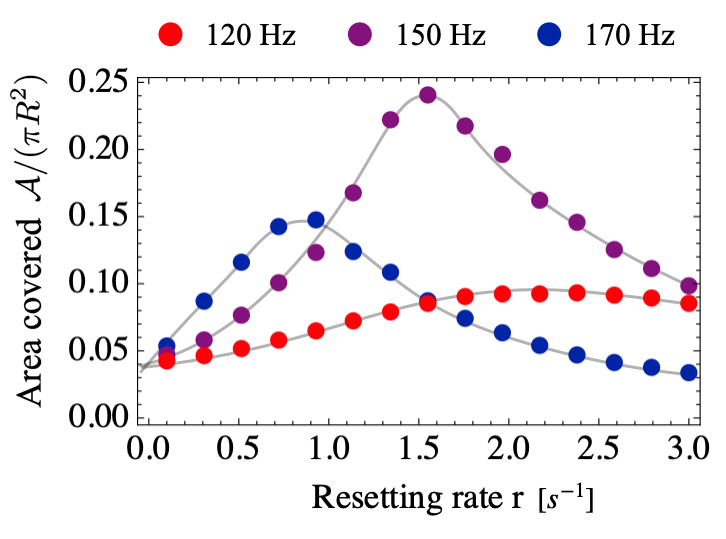}
    \caption{\textbf{Resetting induces optimal covered area.} 
    Covered area $\mathcal{A}$, normalized by the plate area $\pi R^{2}$, as a function of the resetting rate $r$. The analysis is performed for three different shaker frequencies, as reported in the legend, corresponding to different particle parameters (see End Matter, Sec.~\ref{app:experimentaldetails}). Dots represent data points, while the gray solid lines are guides to the eye. Optimal resetting rates are identified as the values of $r$ that maximize the covered area. 
    }
    \label{fig:newexparea}
\end{figure}

\begin{figure*}[t!]
    \centering
    \includegraphics[width = \textwidth]{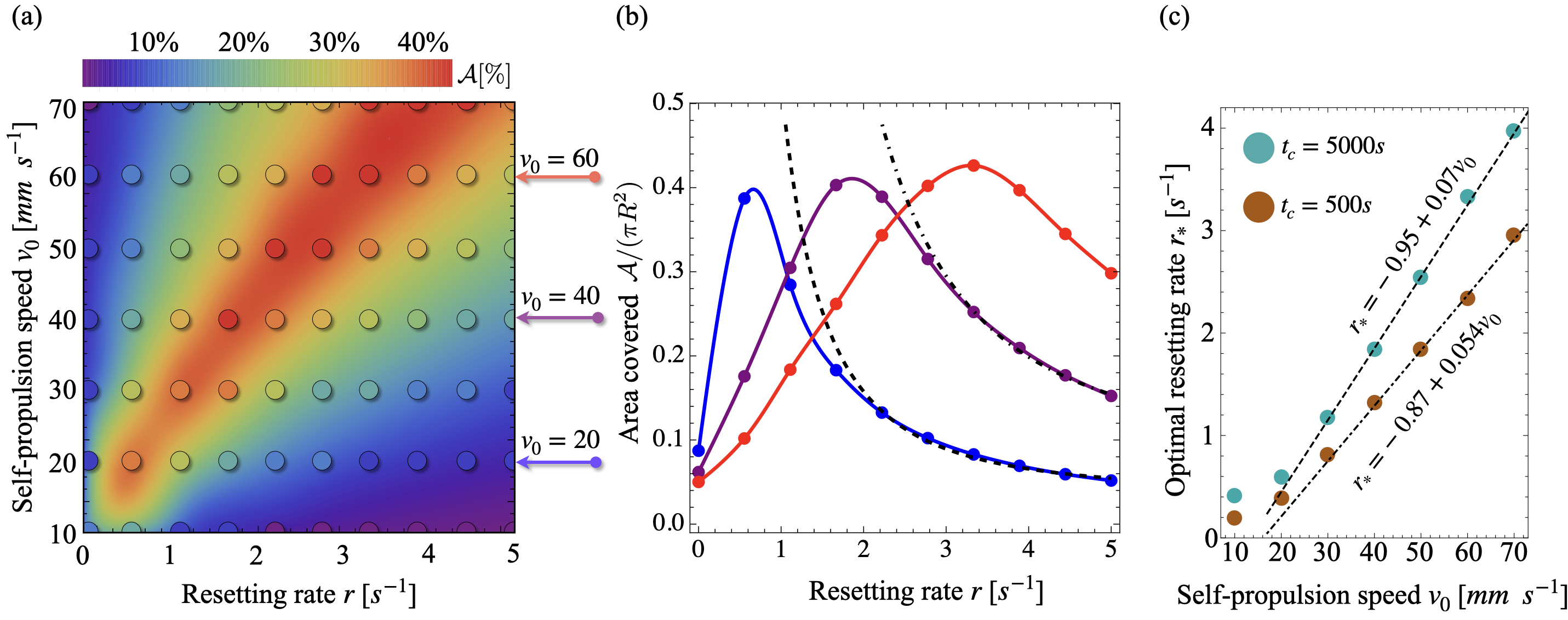}
    \caption{\textbf{Resetting-induced optimal area covered by an active particle.}
    (a) Covered area $\mathcal{A}$ (color gradient) normalized by the area of the circular arena, as a function of the resetting rate $r$ and self-propulsion speed $v_0$. Colored dots represent simulation data; the background is obtained by interpolation. 
    (b) Normalized $\mathcal{A}$ versus $r$ for selected $v_0$ values, indicated by arrows in (a). The two dashed lines denote the predicted $\sim r^{-2}$ scaling at large $r$. The cutoff time is $t_c = 5000\,\text{s}$. 
    (c) Optimal resetting rate as a function of $v_0$, showing the predicted linear behavior (dotted black lines). Blue and red points correspond to cutoff times $t_c = 500\,\text{s}$ and $5000\,\text{s}$, respectively. 
    The remaining simulation parameters are $D_r = 0.85\,\text{s}^{-1}$, $\gamma_T = 10.0\,\text{g}\,\text{s}^{-1}$, and $D_T = 1.59\,\text{mm}^2\,\text{s}^{-1}$. 
    }
    \label{fig:area}
\end{figure*}

\textit{Setup} ---  
We consider an active granular particle~\cite{aranson2007swirling, kudrolli2010concentration, deseigne2010collective, kumar2014flocking, koumakis2016mechanism, baconnier2022selective, lopez2022chirality, scholz2018rotating, antonov2025self}, called a vibrobot~\cite{scholz2018inertial, caprini2024emergent}, moving on a circular plate under the action of an electromagnetic shaker.
A vibrobot is a 3D-printed plastic-made particle consisting of a cylindrical body with a height of~7~mm and a diameter of~$\sigma=15$~mm. Seven legs are attached to the particle body and touch the plate.
The legs are all tilted in the same direction with an angle of 4 degrees so that the translational symmetry of the particle is broken. In this way, when the shaker is on, the vibrobot shows active motion~\cite{caprini2024emergent} which is well-described by the active Brownian particle dynamics with a typical speed~\cite{scholz2018inertial,caprini2024spontaneous}.
Further details on the particle design and the shaker working conditions are reported in End Matter, Sec.~\ref{app:experimentaldetails}.
When the vibrobot hits the container's boundary, the measurement is stopped and the particle is manually placed in the middle of the plate with the same orientation. These trajectories will serve as the basis for the trajectories with stochastic resetting, as explained in detail later.

The dynamics of an active vibrobot are well-described by the equation of motion for a two-dimensional active Brownian particle~\cite{bechinger2016active}. This is experimentally proved in Ref.~\cite{scholz2018inertial}. In our study, we consider shaker conditions such that inertia is small and can be neglected (as reported in Ref.~\cite{caprini2024dynamical}), such that the particle position $\mathbf{x}$ and the orientation $\theta$ evolve as
\begin{subequations}
\label{dynamics}
\begin{align}
&\dot{\boldsymbol{x}}=v_0 \mathbf{n} + \sqrt{2D}\boldsymbol{\xi}\\
&\dot{\theta}=\sqrt{2D_r}\eta \,.
\end{align}    
\end{subequations}
In this dynamics, $D$ is the translational diffusion coefficient, while $\boldsymbol{\xi}$ and $\eta$ are a vector and a scalar delta-correlated Gaussian white noises with zero mean.
The term $v_0$ represents the self-propulsion speed of the vibrobot, with a direction of motion given by $\boldsymbol{n} = [\cos \theta,\sin \theta]$. The orientation angle $\theta$ diffuses with rotational diffusion coefficient $D_r$. Finally, the ratio $v_0/D_r$ defines the persistence length due to the activity, i.e.\ roughly the length run by the particle before its direction of motion changes.
The dynamics~\eqref{dynamics} is complemented by absorbing boundary conditions at the container circular wall, which is placed at a radial position $R$ from the center of the plate.

Resetting events are realized in experiments and simulations by interrupting the particle trajectories after a time $\tau$ extracted from an exponential distribution $p(\tau)\propto \exp(-r \tau )$ characterized by a resetting rate $r$.
 This strategy is implemented in both simulations and experiments. In the latter, the protocol requires manually resetting the particle at the center of the arena only when the vibrobot reaches the boundary and the recording is stopped, rather than at every resetting event. This greatly facilitates the protocol to perform resetting experiments (see End Matter, Sec.~\ref{app:experimentaldetails}).

Here, we focus on calculating the space explored by the particle, referred to as the covered area. After discretizing the space into cells of size approximately equal to the particle radius, the covered area is defined as the fraction of occupied cells. This quantity is computed by tracking the trajectory of the vibrobot’s center of mass with a thickness equal to the particle diameter (see End Matter, Sec.~\ref{sec:discr} for details on the discretization scheme). Consequently, a second visit to the same cell does not increase the covered area, which thus corresponds to the number of ``distinct sites'' in random-walk studies~\cite{donsker1979number,biroli2022number}. We emphasize that defining the covered area based on the particle size provides only a lower bound on the total area explored, as many realistic systems may have a larger radius of exploration.

Specifically, we focus experimentally and numerically on the mean covered area $\mathcal{A}$ explored up to the time $\tau_S$ required for the particle to reach the boundary. If the particle is reset in the middle of the container at a time $\tau<\tau_S$ due to a random resetting event, the calculation of $\mathcal{A}$ continues, since the particle has not yet touched the boundary. 
However, we include an upper cut-off time $t_C$ at which the measurement is stopped even if the container's boundary is not reached. This mimics the desire to achieve a goal in finite time, and excludes unfavorable exploration strategies that are too slow.  
Hence, we are interested in the mean area explored up to a time $\min\{\tau_S,t_C\}$.

\textit{Results} --- 
By visual inspection of the particle trajectory (Fig.~\ref{fig:paths}~(b)), we observe that both experiments and simulations qualitatively predict a non-monotonic behavior with the resetting rate $r$ for the area covered by the particle.
 To quantify this observation, we study $\mathcal{A}$ as a function of $r$ for three experimental configurations corresponding to different self-propelled speeds, $v_0=11.4,26.1,63.3$ mm $\text{s}^{-1}$ (Fig.~\ref{fig:newexparea}). Experimentally, different self-propulsion speeds are obtained by varying the shaker's frequency (see End Matter, Sec.~\ref{app:experimentaldetails}). The case $r=0$ corresponds to the absence of resetting, while the maximal, statistically relevant, resetting rate shown corresponds to 3 Hz.
The mean covered area displays a non-monotonic behavior with $r$: it increases until a maximum (up to 30\%) and afterward decreases again for further increasing values of $r$.

Simulations allow us to more systematically explore the role of the particle speed $v_0$ and resetting rate $r$ (Fig.~\ref{fig:area}~a). In the heat map, bright and dark colors correspond to the higher and lower values of the mean covered area $\mathcal{A}$ (up to a covered area fracton of 40\%).
In the absence of resetting, $\mathcal{A}$ decreases monotonically with the self-propulsion speed. Indeed, as the persistence length $v_{0}/D_{r}$ becomes large compared to the plate radius, the particle moves ballistically toward the container boundary, exploring a narrow region of the domain.

In the presence of resetting, a competition arises. Resetting generates an effective attraction toward the origin, which forces the particle to repeatedly explore the region near the center of the plate but also allows it to explore the arena for longer times by avoiding the absorbing walls. This leads to a non-monotonic dependence of the covered area $\mathcal{A}$ on $r$ (Fig.~\ref{fig:area}~b), a behavior observed systematically across several values of the self-propulsion speed (Fig.~\ref{fig:area}~a). 
The cutoff time $t_{C}$ does not qualitatively alter the results, preserving the non-monotonic behavior of the covered area (see End Matter, Sec.~\ref{app:cut}).
In addition, this non-monotonicity also occurs for other resetting-time distributions, as verified for gamma-distributed resetting times (see End Matter, Sec.~\ref{app:cut}), demonstrating the generality of the observed phenomenon.

To gain analytical insights into the experimentally and numerically observed covered area, we consider the regime where the particle's persistence length $v_0/D_r$ is large compared to the system size. 
Consequently, except when interrupted by resetting events, the particle moves ballistically toward the boundary with speed $v_0$.
This allows us to extract the behavior at small and large $r$, as well as an estimate for the optimal resetting rate $r^*$. 
At small resetting rates, the exploration process is dominated by a small number $n$ of independent trajectories, each covering an area $A_1$. 
Since the process ends at first passage to the boundary, the term $n \approx r \langle \tau_S \rangle$, where $\langle \tau_S \rangle$ is the mean first-passage time. 
For a ballistic trajectory, $A_1 = 2 \sigma v_0 / r$, using the typical length $v_0/r$. 
Because the first-passage problem in the high-persistence regime is rotationally symmetric, it can be reduced to a one-dimensional problem along a ray from the origin to the boundary. 
The mean first-passage time satisfies 
$\mathcal{L}_\text{HP}^\dagger(x) \langle \tau_S \rangle(x) - r \langle \tau_S \rangle(x) + r \langle \tau_S \rangle(x_R) = -1$, 
with $\langle \tau_S \rangle(R) = 0$, where $x$ parametrizes the ray and $x_R = 0$ is the resetting location at the origin. 
Here, $\mathcal{L}_\text{HP}^\dagger(x)$ is the adjoint of the generator of high-persistence trajectories, which takes the simple form $\mathcal{L}_\text{HP}^\dagger(x) = v_0 \partial_x$. 
The solution of this differential equation is $\langle \tau_S \rangle = (e^{r R/v_0} - 1)/r$. 
Combining these results, we then expect at small resetting rates
\begin{equation}
    \mathcal{A} \approx 2 \sigma v_0 \langle\tau_S \rangle = 2 \sigma R+ \frac{\sigma R^2 }{v_0}r + \frac{\sigma R^3}{3 v_0^2} r^2 +...
\end{equation}
When no resetting takes place, the covered area coincides with that of a straight like from the origin to the boundary, $2 \sigma R$, while resetting increases this area with a linear leading-order correction. 

At large resetting rates, we can no longer assume that the areas are independent as in the small-$r$ limit, since overlaps become significant. In this regime, rapid resetting generates a densely explored region near the origin, with an effective radius given by the typical distance a particle travels before a reset, $v_0/r$. 
Hence, we expect that the particle covers a dense region with area $\mathcal{A} \approx (v_0/r)^2$. 
Moreover, the exploration process terminates once a rare event occurs, allowing the particle to reach the boundary before a resetting event. Consequently, we expect the covered area to scale as
\begin{equation}
    \mathcal{A} \approx \pi (v_0/r)^2 + 2 \sigma R \,,
\end{equation}
for large resetting rates.

To estimate the optimal resetting rate $r_*$, we equate the leading scaling behaviors at small and large rates, 
$\pi \frac{v_0^2}{r_*^2} = \frac{\sigma R^2 r_*}{v_0}$. 
This yields a simple linear relation between the optimal resetting rate and the self-propulsion speed, $r_* \sim v_0$:
\begin{equation}
    r_* \sim \frac{v_0}{R^{2/3}\sigma^{1/3}}\,.
\end{equation}
This behavior is confirmed in Fig.~\ref{fig:area}(c) where it is observed even beyond the large-persistence regime. 
This can be explained through Cavalieri's principle, which states that the area of a weakly curved trajectory (i.e., large but finite persistence) is equal to that of its straight, ballistic counterpart, as shown in End Matter (Sec.~\ref{app:area}). 
As long as self-intersections are sufficiently rare, the high-persistence scaling results are expected to hold. 
Furthermore, the scaling of the optimal resetting rate, $r_* \sim v_0$, implies that the area at optimality, $\mathcal{A}_*$, is approximately independent of the self-propulsion speed. 
Since resetting typically incurs a cost, e.g., fuel consumed when returning to a central location, this suggests that favorable strategies may correspond to lower degrees of activity in this regime. 
The approximate independence of the covered area at optimality with respect to activity is confirmed in our analysis (Fig.~\ref{fig:area}(b)), although we observe a slight increase in $\mathcal{A}_*$ with $v_0$. 
This indicates that, beyond the high-persistence regime, particles with higher activity are more effective reporters of the environment.

\textit{Discussion } ---  We have demonstrated an optimal strategy to maximize spatial exploration in active matter systems, a problem of broad relevance to search processes in living organisms, synthetic microswimmers, and macroscopic robots. Our results show that active particles benefit from intermittently restarting their exploration from a central location. This effect is quantified by an activity-dependent optimal resetting rate that maximizes the covered area. Our findings are demonstrated experimentally using active granular particles (vibrobots) and confirmed by active Brownian particle simulations, which systematically establish the robustness of the results by varying the activity beyond the experimental conditions. 
The simulations further clarify that our results originate solely from the competition between persistence dynamics and the resetting rate. 
Other forces typically governing the vibrobot dynamics, such as self-alignment and dry friction, are not essential for our findings. 
Importantly, we discover that the optimal resetting rate scales linearly with the self-propulsion speed, such that the characteristic length scale $v_0/r$ remains constant at optimality. This scaling relation provides a simple yet powerful design principle for efficient search in active systems. 

Our findings connect a predominantly theoretical topic in statistical mechanics, such as resetting, to applications involving granular robots, including efficient spatial exploration and resource acquisition. We anticipate that this study will stimulate further research, from theoretical efforts aimed at rationalizing our results within frameworks based on the Feynman–Kac formalism, stochastic functionals or large deviation theories~\cite{agranov2020airy,majumdar2020statistics,majumdar2007brownian}, to the development of more efficient strategies employing different resetting-time distributions -- for instance, see End Matter, Sec.~\ref{app:cut}, for a discussion of gamma-distributed resetting times. The experimental strategy introduced here to investigate resetting overcomes limitations due to insufficient statistics, a common challenge in experimental studies. This approach can also be applied to colloidal systems or more complex biological organisms, where efficient spatial exploration underlies critical processes such as resource acquisition, avoidance of harmful conditions, and surface colonization for biofilm formation.

\vskip0.2cm
\noindent
\textit{Acknowledgements} ---
LC acknowledges the European Union MSCA-IF fellowship for funding the project CHIAGRAM. KSO acknowledges support from the Alexander von Humboldt Foundation.
HL and KO acknowledges support by the Deutsche Forschungsgemeinschaft (DFG) through the grant number LO 418/29-1.

\vskip0.2cm
\noindent
\textit{Author Contributions} ---
KO, HL, and LC design and conceptualize the research.
LC designed and performed the experiment. KO performed numerical simulations.
All the authors equally contribute to the manuscript writing.

\bibliographystyle{apsrev4-2}
\bibliography{OlsenEtAl.bib}

\newpage
\section*{End Matter}

\subsection{Details of the experiment}
\label{app:experimentaldetails}

\begin{figure}
    \centering
    \includegraphics[width=6.5cm]{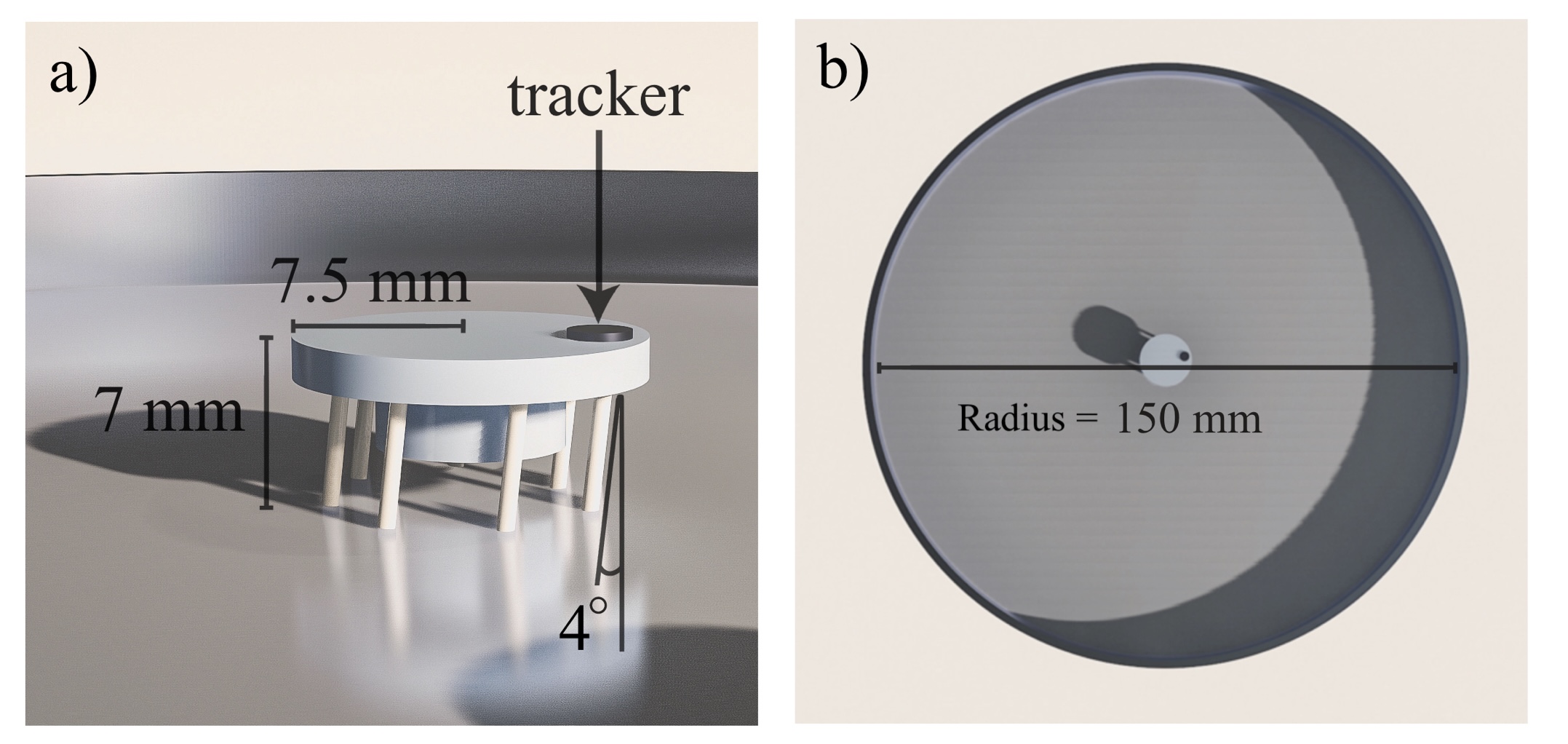}
    \caption{{\bf Vibrobot and setup illustration.} (a) Schematic side-view of the vibrobot.
    (b) Illustration of the setup top-view of a particle moving on the vibrating plate.}
    \label{fig:setup}
\end{figure}

\paragraph*{Vibrobot design --} 
The active vibrobot is fabricated from a proprietary photo-polymer by stereolithographic 3D printing and consists of a cylindrical core, a cap, and seven legs attached to the cap. 
The core has a diameter of $9\,\text{mm}$ and a height of $4\,\text{mm}$, whereas the cap placed on top of the core has a diameter of $15\,\text{mm}$ and a height of $2\,\text{mm}$. 
Seven cylindrical legs with a diameter of $0.8\,\text{mm}$ and a length of $5\,\text{mm}$ are attached to the cap and arranged in a regular heptagon around the core. The resulting particle has a total height of $7\,\text{mm}$, and covers a circular area on the plate with a diameter of 15~mm (Fig.~\ref{fig:setup}(a)-(b)). All legs are tilted by $4^\circ$ compared to the vertical in the same direction, breaking the translational symmetry of the body and giving rise to active motion. 
The vibrobot has a mass $m = 0.83\,\text{g}$ and an approximate moment of inertia $J = 17.9\,\text{g}\,\text{mm}^2$. A black sticker with a white circle is placed on the particle cap, aligned with the direction of the leg tilt.


\noindent
\paragraph*{Experimental setup --}
The vibrobot shows active motion when placed on a vertically vibrating circular plate. The plate has a diameter of $300\,\text{mm}$ and a height of $15\,\text{mm}$ and is positioned horizontally compared to the ground. Vertical vibrations are generated by an electromagnetic shaker, with frequency $f$ and amplitude $A$, connected to the plate. A plastic barrier surrounding the plate prevents the vibrobot from falling off.


\paragraph*{Data acquisition method --}
Data are acquired by recording images with a high-speed camera operating at a temporal resolution of 50 frames/s and a spatial resolution of $0.3\,\text{mm/pixel}$. Particle positions are determined from the images using a standard tracking algorithm with sub-pixel accuracy, while particle orientation is obtained by tracking the position of the white circle relative to the center of mass. Translational and angular velocities are then computed as $\mathbf{v} = (\mathbf{x}(t+\Delta t) - \mathbf{x}(t))/\Delta t$ and $\omega = (\theta(t+\Delta t)-\theta(t))/\Delta t$, respectively, with $\Delta t = 0.02\,\text{s}$ corresponding to the acquisition rate of 50 frames/s.


\paragraph*{Parameter extraction for active vibrobots --}
As shown in Ref.~\cite{scholz2018inertial}, the vibrobot can be described as an underdamped active Brownian particle with velocity $\mathbf{v}=\dot{\mathbf{x}}$ and angular velocity $\omega=\dot{\theta}$, namely,
%
\begin{subequations}
\label{eq:app_dynamics}
\begin{align}
    \dot{\bm{v}}(t) &= -\gamma_t \bm{v}(t) + \gamma_t v_0 \bm{n} (t)+ \gamma_T \sqrt{2 D_t} \bm{\xi}(t)\\
    \dot \omega(t) &= -\gamma_R \omega(t) + \gamma_r \sqrt{2 D_r} \zeta(t) \,,
\end{align}
\end{subequations}
where $m$ and $J$ denote the particle mass and moment of inertia, respectively. The coefficients $\gamma_t$ and $\gamma_r$ represent the translational and rotational friction, while $D_t$ and $D_r$ are the translational and rotational diffusion constants. The term $\gamma_t v_0 \mathbf{n}$ accounts for the active force driving the motion, with $v_0$ the characteristic running velocity of the vibrobot and $\mathbf{n} = (\cos\theta(t),\,\sin\theta(t))$ its orientation, i.e., the direction along which the legs are tilted.


To estimate the five free parameters of dynamics~\eqref{eq:app_dynamics}, we match five observables between experiments and simulations. In particular, we compute the velocity distribution $P(\mathbf{v})$, the angular velocity distribution $P(\omega)$, the mean-square displacement $\langle [\mathbf{x}(t)-\mathbf{x}(0)]^{2}\rangle$, the angular mean-square displacement $\langle [\theta(t)-\theta(0)]^{2}\rangle$, and the cross-correlation between velocity and orientation 
$\langle \mathbf{v}(t)\mathbf{n}(0)\rangle - \langle \mathbf{n}(t)\mathbf{v}(0)\rangle$. These observables are evaluated from both experimental data and numerical simulations initiated with a trial set of parameters. The total deviation between experimental and simulated results is then computed and minimized iteratively using a Nelder–Mead optimization scheme.

Compared to Ref.~\cite{scholz2018inertial}, here we consider three different shaker conditions of frequency $f$ and amplitude $A$, which lead to different sets of vibrobot parameters:
\begin{center}
\begin{tabular}{ l | c | c  | c }
 f [Hz] & 120 & 150 & 170 \\ 
 $v_0$ [mm \,s$^{-1}$] & 63.6  & 26.1 & 11.4 \\  
 $D_t$ [mm$^2$ s$^{-1}$] & 1.59 & 0.57 & 8.56 \\
 $D_r$ [s$^{-1}$] & 0.85 & 0.29 & 0.11 \\
 $\gamma_t$ [g\,s$^{-1}$] & 10.0 & 75.8& 281.5 \\
 $\gamma_R$ [g\, mm$^2$ s$^{-1}$]& 305 & 325 & 311 .\\
\end{tabular}
\end{center}
In all three cases, the inertial time $m/\gamma_t$ and the rotational inertial time $J/\gamma_r$ are significantly shorter than the persistence time $1/D_r$. Consequently, the evolutions of the velocity $\mathbf{v}$ and angular velocity $\omega$ can be neglected, leading to overdamped equations of motion for the position $\mathbf{x}$ and the orientation angle $\theta$. 
This simplification reduces the parameter space to be explored in the numerical study. Specifically, since the particle size and the system size are fixed, the only relevant parameter is the persistence length $v_0/D_r$, as discussed in the main text.

\paragraph*{Resetting implementation --} 
Our resetting experiment follows a three-step protocol. (i) A particle is manually placed at the center of the container with a random orientation. (ii) We record its trajectory until it reaches the container boundary, in a time $\tau_S$. Steps (i) and (ii) are repeated to obtain $N$ trajectories, where the particle reaches the boundary after a time $\tau_S$ for each trajectory. 
iii) We extract a time $\tau$ from an exponential distribution $p_r(\tau)$ with rate $r$ and compare with the first measured value for $\tau_S$.
The entire trajectory is retained if $\tau>\tau_S$, i.e., the particle reaches the boundary before the resetting time.
Otherwise, if $\tau<\tau_S$, the trajectory is truncated at time $\tau$, meaning the particle is reset before reaching the boundary. In this case, after the reset, the particle trajectory continues by concatenating it with the next trajectory. 
This process is repeated until the particle reaches the boundary (so that the extracted $\tau$ is larger than the corresponding measurement for $\tau_S$). 
If the time accumulated from these trajectories is larger than $t_C$, the process is stopped because the particle has not reached the boundary within the cutoff time.
Our experimental strategy offers several advantages. First, it allows us to use the same set of $N$ trajectories to compute any observable for different resetting rates. Second, it reduces experimental error, as the resetting time is imposed \textit{a posteriori} rather than manually in each experiment.

\subsection{Discretization scheme for area fraction} 
\label{sec:discr}
\begin{figure}[t]
    \centering
    \includegraphics[width = 6.0cm]{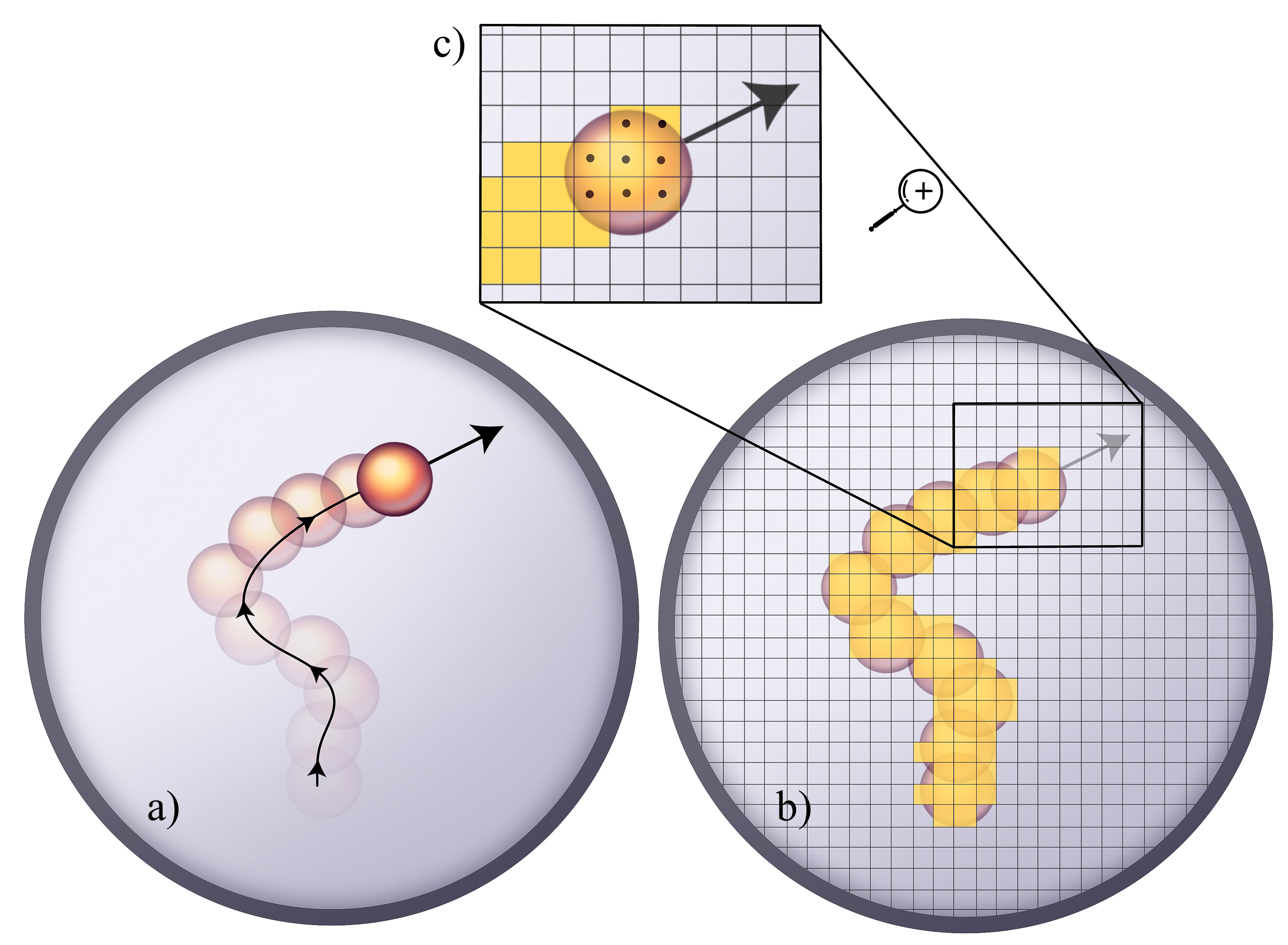}
    \caption{Discretized particle path and area, with $\delta = 0.133 r_p$ and $\mathcal{A}/(\pi R^2)\approx 0.1$. Unoccupied and occupied cells are shown in gray and yellow. Overlapping black path shows center of mass trajectories.}
    \label{fig:discr1}
\end{figure}

To estimate the covered area $\mathcal{A}/(\pi R^2)$, we discretize space into cells of linear extent $\delta$ (see Fig.~\ref{fig:discr1} for an illustration). 
At time $t$, the particle has a center-of-mass coordinate $(x_t,y_t)$, which upon rounding to the nearest cells translates into a cell coordinate $(x_t/\delta,y_t/\delta)$. 
The cells covered by the particle at that instant are those with indices $(n,m)$ for which
\begin{equation}
     \left(\frac{x_t}{\delta}+n\right)^2 + \left(\frac{y_t}{\delta}+m\right)^2 \leq \left( \frac{r_p}{\delta}   \right)^2 \,,
\end{equation}
where $r_p$ coincides with the particle radius.
This defines, up to time $t$, a number of occupied $n_O(t)$ and unoccupied cells $n_U(t)$. The normalized covered area is obtained as 
\begin{equation}
    \frac{\mathcal{A}}{\pi R^2} = \frac{n_O(t)}{n_O(t) + n_U(t)} \,.
\end{equation}
Since the rounding error is not crucial for the results explored here, we pick $\delta = 0.4 r_p$.

\begin{figure}[t!]
    \centering
    \includegraphics[width=8.65cm]{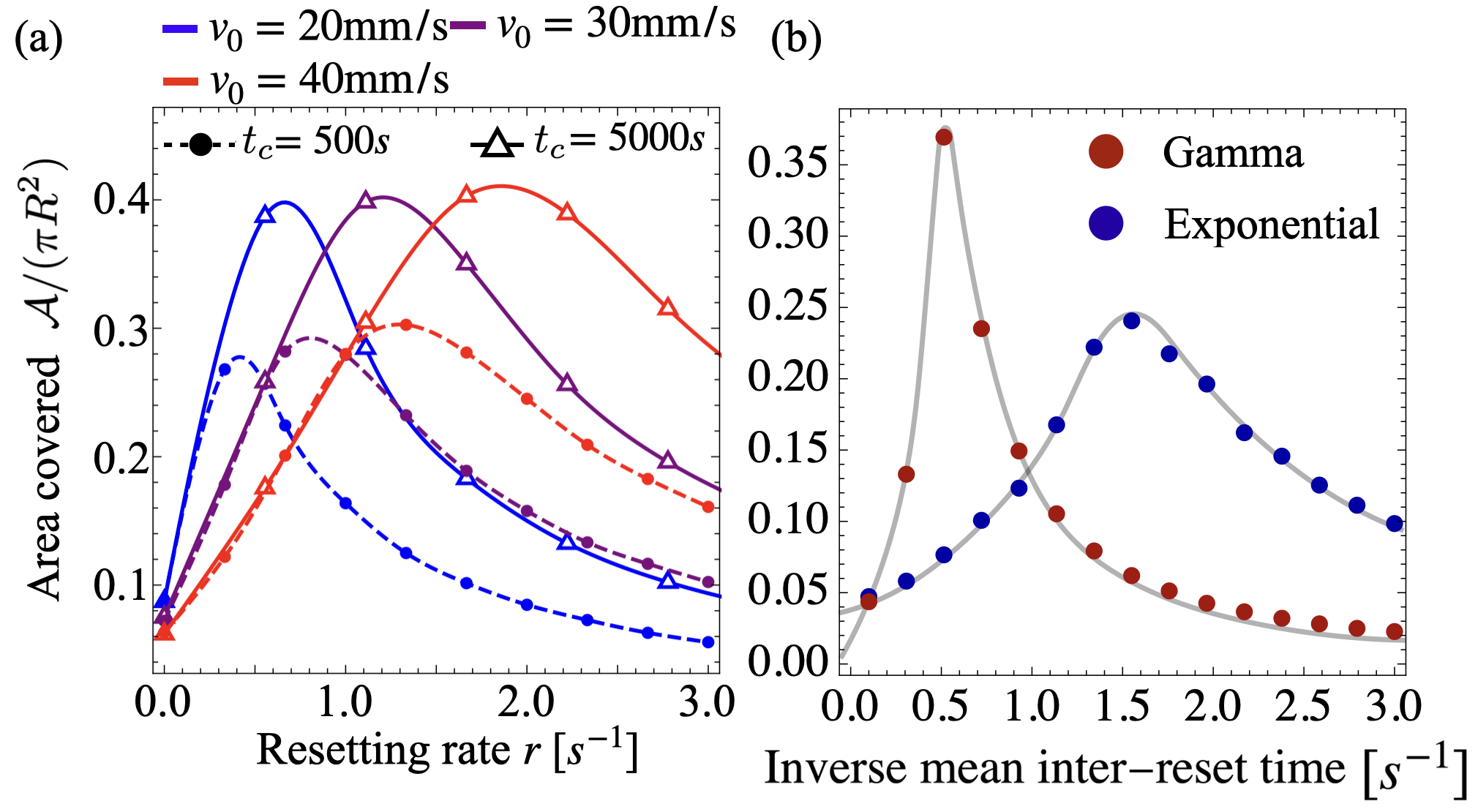}
    \caption{ 
    (a) Covered area $\mathcal{A}$ normalized by the plate area obtained for two temporal cutoffs values, $t_c$.  
    (b) Normalized $\mathcal{A}$ calculated for exponential, $p(\tau) = r \exp(- r \tau)$, and gamma distributions, $p(\tau) = \tau^{k-1} \exp(-\tau/\theta)/\Gamma(k)$, with identical mean and $k=10$. 
    (a) and (b) are obtained from simulations and experiments, at shaker frequency of $150$ Hz.
    }
    \label{fig:combined}
\end{figure}

\subsection{Temporal cutoff and non-Poissonian strategies}\label{app:cut}
{\paragraph*{Cutoff time role} --}
In our protocol, a temporal cutoff is applied to exclude excessively long exploration strategies, mimicking tasks such as drones, robots, or animals searching for targets or mapping an area within a finite time. The covered area $\mathcal{A}$ is measured for two cutoff times $t_C$, differing by an order of magnitude (Fig.~\ref{fig:combined}~(a)). 
The change in $t_C$ does not alter the qualitative trend of $\mathcal{A}$ as a function of the resetting time $r$.

{\paragraph*{Non-Poissonian strategies} --}
Exponential resetting times are not always optimal for minimizing first-passage times in passive Brownian particles~\cite{evans2025stochastic}, or covered areas.
This raises the question of whether non-exponential strategies may be relevant to the area coverage problem studied here. 
From our experimental data, we report the area covering fraction for both exponential, $p(\tau) = r \exp(- r \tau)$, and gamma-distributed, $p(\tau) = \tau^{k-1} \exp(-\tau/\theta)/\Gamma(k)$, inter-reset times (Fig.~\ref{fig:combined}~(b)). 
This analysis suggests that sharper resetting strategies may further enhance area exploration in active particles.

\subsection{Correction to high-persistence regime}\label{app:area}
Here, we show that the area of straight ribbons in two dimensions remains unchanged when small curvatures are included. Let $\gamma(s)$ be a planar curve with internal coordinate $s\in(0,L)$ and denote the normal vector at a point $s$ as $N(s)$, such that the tangent vector is $\dot \gamma(s) = T(s)$. The signed curvature $\kappa(s)$ at $s$ is defined by $\dot T(s) = \kappa(s) N(s)$. By considering a parallel curve $\tilde \gamma(s) = \gamma(s) + u N(s)$ obtained by displacing a distance $u$ along the local normal, the tangent to the parallel curve is $\tilde T(s) = T(s) + u \dot N(s)$. Using the planar Frenet-Serret relation $\dot N(s) = -\kappa(s) T(s)$, we have $\tilde T(s) = (1 - u \kappa(s)) T(s)$. The area element of a small parallelepiped spanned by increments $ds$ and $du$ along the $s$ and $u$ directions is $dA = (1 - u \kappa(s))\, ds\, du$, so that the total area of the strip is
\begin{equation}
    \mathcal{A} = \int_{-w/2}^{w/2} du \int_0^L ds (1-\kappa(s) u)=L w \,.
\end{equation}
where the term proportional to $u$ immediately cancels due to the antisymmetry of the integral and $\mathcal{A}$ is the same as if the strip had no curvature. For active particles, the typical paths before resetting have length $v_0/r$ and width $2\sigma$. By the above argument, the high-persistence approximation is expected to hold well as long as path self-intersections are rare, even if the trajectories exhibit curvature due to finite persistence.

\end{document}